\documentclass[useAMS,usenatbib]{mn2e}

\usepackage{amssymb}
\usepackage{graphicx}
\usepackage{mathptmx}
\usepackage{amsmath}
\usepackage{color}

\title[Molecular cloud and star cluster mass functions]{From the molecular-cloud- to the embedded-cluster-mass function  \\
with a density threshold for star formation}

\author[G. Parmentier]
{Genevi\`eve Parmentier$^{1}$\thanks{Humboldt Fellow - E-mail: gparm@astro.uni-bonn.de} \\ 
$^{1}$Argelander-Institut f\"ur Astronomie, Bonn Universit\"at, Auf dem H\"ugel 71, D-53121 Bonn, Germany}

\begin{document}

\date{Accepted 2010 December 24.  Received 2010 December 15; in original form 2010 September 28
}
\pagerange{\pageref{firstpage}--\pageref{lastpage}} \pubyear{2011}

\maketitle

\label{firstpage}

\begin{abstract}

The mass function $dN \propto m^{-\beta_0}dm$ of molecular clouds and clumps is 
shallower than the mass function $dN \propto m^{-\beta_\star}dm$ of young star clusters, 
gas-embedded and gas-free alike, as their respective mass function indices are 
$\beta_0 \simeq 1.7$ and $\beta_\star \simeq 2$.  
We demonstrate that such a difference can arise from different mass-radius
relations for the embedded-clusters and the molecular clouds (clumps) hosting them.
In particular, the formation of star clusters with a constant mean {\it volume}
density in the central regions of molecular clouds of constant mean 
{\it surface} density steepens the mass function from clouds to embedded-clusters.  
This model is observationally supported since the mean surface density of molecular clouds 
is approximately constant, while there is a growing body of evidence, in both Galactic 
and extragalactic environments, that efficient star-formation requires a hydrogen
molecule number density threshold of $n_{th} \simeq 10^{4-5}\,cm^{-3}$.    

Adopting power-law volume density profiles of index $p$ for spherically symmetric 
molecular clouds (clumps), we define two zones within each cloud (clump): a central 
cluster-forming region, 
actively forming stars by virtue of a local number density higher than $n_{th}$, and an 
outer envelope inert in terms of star formation.  
We map how much the slope of the cluster-forming region mass function differs from that 
of their host-clouds (clumps) as a function of their respective mass-radius relations
and of the cloud (clump) density index.  We find that for constant surface density clouds 
with density index $p \simeq 1.9$, a cloud mass function of index $\beta_0 = 1.7$ gives 
rise to a cluster-forming region mass function of index $\beta \simeq 2$.
Our model equates with defining two distinct SFEs:
a global mass-varying SFE averaged over the whole cloud (clump), and a local 
mass-independent SFE measured over the central cluster-forming region.  While the global
SFE relates the mass function of clouds to that of embedded-clusters, the local SFE 
rules cluster evolution after residual star-forming gas expulsion.  That the cluster mass
function slope does not change through early cluster evolution implies a mass-independent
local SFE and, thus, the same mass function index for cluster-forming regions 
and embedded-clusters, that is, $\beta = \beta_\star$.  Our model
can therefore reproduce the observed cluster mass function index $\beta_\star \simeq 2$.
      
For the same model parameters, the radius distribution also steepens from 
clouds (clumps) to embedded-clusters, which contributes to explaining observed
cluster radius distributions.

\end{abstract}


\section{Introduction}
\label{sec:intro}
The star formation efficiency (SFE) achieved by star cluster gaseous precursors at the onset of residual star-forming gas expulsion is a crucial quantity since it influences the cluster dynamical response to gas expulsion significantly \citep[the so-called violent relaxation;][]{hil80,gey01,bau07,pro09}.  Specifically, the SFE is tightly related to whether the cluster survives violent relaxation and, if it survives, what mass fraction of its stars it retains.  The SFE being the ratio between the stellar mass of embedded-clusters and the initial gas mass of their precursor molecular clouds, the comparison of the mass functions of young star clusters and of molecular clouds holds the potential of highlighting whether the SFE varies with molecular cloud mass.       

The mass function $dN \propto m^{-\beta_0}dm$ of giant molecular clouds (GMCs) in the Local Group of galaxies has an index $\beta_0 \simeq 1.6$-$1.7$ \citep{ros05, bli06} \citep[see also ][for the case of the GMC mass function in the Magellanic Clouds]{fuk08}.  These GMCs, when compressed by the high pressure of violent star-forming environments, are expected to be the parent clouds of massive star clusters forming profusely in galaxy mergers and starbursts \citep{jog92,jog96}.  The same slope $\beta_0 \simeq 1.6$-$1.7$ is also found for the mass function of density enhancements contained by GMCs -- referred to as molecular clumps \citep{lad91,kra98,won08}.  In quiescent disc galaxies such as the Milky Way, those are observed to be the progenitors of open clusters \citep[][and references therein]{hp94,ll03}.  

In contrast to molecular clouds and clumps, the mass function $dN \propto m^{-\beta_{\star}}dm$ of embedded and young clusters is, in most cases, reported to have an index $\beta_{\star} \simeq 2$ \citep[e.g.][]{zf99,bik03,ll03,oey04}, which is steeper than the mass function of molecular structures.   
Given the uncertainties affecting both slopes, the significance of the $\beta_{\star} - \beta_0$ difference remains uncertain.  \citet{elm96} suggest that error-free measurements of GMC masses may bring the mass function slopes of young star clusters and GMCs in agreement.  Conversely, one can consider that the slope difference is significant, which is the approach we adopt in this paper.

The question we set to answer is: what process of the physics of cluster-formation steepens the power-law mass function of molecular clouds and clumps from $\beta_0=1.7$ to $\beta_{\star}=2$?  The $\beta_{\star}-\beta_0$ difference suggests that the SFE is a decreasing function of the cloud (clump) mass.  Besides sounding counter-intuitive, a mass-varying SFE is necessarily conducive to mass-dependent cluster infant weight-loss since the fraction of stars remaining bound to clusters through violent relaxation is a sensitive function of the SFE \citep[e.g. fig.~1 in][]{par07}.  This does not seem to be supported by observations of young star clusters, as their mass function slope is reported to remain invariant with time over their first 100\,Myr of evolution \citep {ken89,mck97,ll03,zf99,oey04,dow08,cha10}.  

However, this contradiction is apparent only for it is worth keeping in mind that the SFE driving cluster violent relaxation is the mass fraction of gas turned into stars {\it over the volume of gas forming stars}.  And this volume of star-forming gas may not coincide with the entire cloud (clump).  In what follows, we refer to it as the {\it cluster-forming region (CFRg)}.  Its SFE is the {\it local} SFE and its mass function slope is $-\beta$.  The invariance of the young cluster mass function slope at early time suggested by many observations demands a mass-independent {\it local} SFE.  That is, the mass fraction of gas turned into stars by a CFRg is independent of its mass.  This in turn implies that the slopes of the CFRg and embedded-cluster mass functions are identical: $\beta = \beta_{\star}$.  Therefore, understanding the slope difference $\beta_\star - \beta_0$ between the star cluster and molecular cloud (clump) mass functions equates with understanding why the CFRg mass function is steeper than the mass function of their host clouds (clumps), i.e. $\beta = \beta_\star \simeq 2$ and $\beta_0 \simeq 1.7$.  The $\beta - \beta_0$ difference suggests that the mass fraction of star-forming gas inside molecular clouds (clumps) is a decreasing function of the cloud (clump) mass.  Besides, that the CFRg represents a fraction only of its host cloud (clump) allows us to define a {\it global} SFE, namely, the ratio between the mass in stars formed inside a molecular cloud (clump) and its initial gas mass.  The {\it global} SFE is relevant to explaining the $\beta_{\star}-\beta_0$ slope difference, but irrelevant for modelling cluster violent relaxation.  

What could be the origin of a mass-varying mass fraction of star-forming gas inside molecular clouds (clumps)?
In other words, why should the {\it global} SFE vary with the cloud (clump) mass such that $\beta_\star \neq \beta_0$?    

The mean {\it surface} density of GMCs in our Galaxy is about constant \citep[fig.~8 in][]{bli06}.  This result is reminiscent of Larson's seminal study \citep{lar81} showing that molecular clouds have approximately constant mean column densities \citep[see also][]{lom10}.  On the other hand, star-forming regions are observed to be systematically associated with dense molecular gas, namely, with {\it number} densities of at least $n_{H2} \simeq 10^4$-$10^5\,cm^{-3}$ \citep[][]{mue02,gao04,fau04,fon05,shi03,wu10}.  See also fig.~1 and section 3 in \citet{par10} for a discussion.  Several studies have therefore suggested that star formation requires a gas {\it volume} (or {\it number}) density threshold \citep[e.g.][]{eva08,wu10,lad10}.     

\begin{figure}
\includegraphics[width=\linewidth]{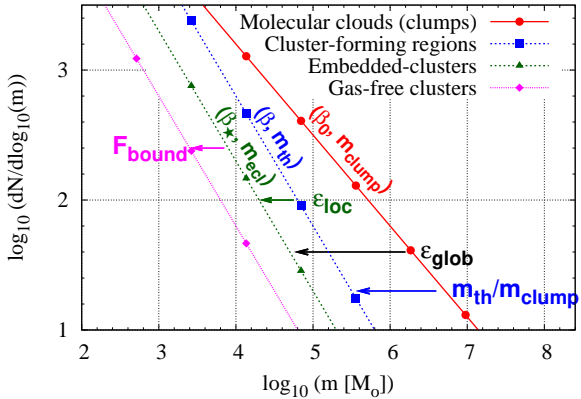}
\caption{Illustration of the different mass functions tackled through the paper.  From right to left: the clump (cloud) mass function $dN \propto m_{clump}^{-\beta_0}\,dm_{clump}$, the CFRg mass function $dN \propto m_{th}^{-\beta}\,dm_{th}$, the embedded-cluster mass function $dN \propto m_{ecl}^{-\beta_{\star}}\,dm_{ecl}$.  For the sake of completeness, the cluster mass function at the end of violent relaxation is also shown as the leftmost straigthline.  The horizontal arrows depict mass ratios relating pairs of mass functions.  From right to left: the mass fraction $m_{th}/m_{clump}$ relates the CFRg and clump mass functions, the global SFE $\epsilon_{global} = m_{ecl}/m_{clump}$ quantifies the embedded-cluster stellar mass contained within molecular clouds (clumps), the local SFE $\epsilon_{loc} = m_{ecl}/m_{th}$ is the CFRg mass fraction turned into stars, the bound fraction $F_{bound}$ quantifies the mass fraction of stars which stays bound to a cluster when violent relaxation is over (i.e. after infant weight-loss).  Vertical scaling is arbitrary. \label{fig:skMF} }
\end{figure}

In this contribution, we develop a model for a spherically symmetric molecular cloud (clump) with a power-law density profile which forms a star cluster in its central region.  We demonstrate that if the mass-radius relation of CFRgs differs from that of the host-clouds (clumps), then the slopes of their respective mass functions are different too (i.e. $\beta \neq \beta_0$).  This will be the case for molecular clouds of constant mean surface density hosting CFRgs of constant mean volume density.  Applying the same model to the radius distribution, we will show that it can also contribute to explaining why the distribution of star cluster half-light radii is significantly steeper than the  distribution of GMC sizes.  Figure \ref{fig:skMF} summarises the different mass functions encompassed through the paper, along with their respective index and the various mass ratios relating them. 

Note that this paper does not intend to explain the slope of the {\it stellar} initial mass function.  That issue is addressed in \citet{sha10} whose model successfully reproduces the Salpeter slope of $-2.35$ for a population of pre-stellar cores exceeding a volume density threshold in a fractal cloud.

The outline of the paper is as follows.  Section \ref{sec:evid} summarises two types of evidence supporting the hypothesis of a constant mean {\it volume} density for CFRgs.  One hinges on the early dynamical evolution of star clusters.  The second is based on the mapping of star-forming regions with dense molecular gas tracers.  In Section \ref{sec:cc}, we build a model relating the power-law mass function of CFRgs to the power-law mass function of their host clouds (clumps).  We map how the slope difference $\beta - \beta_0$ varies as a function of the mass-radius relation and density profile of molecular clouds (clumps).  In Section \ref{sec:implic}, we discuss the implications of our model.  Specifically, we focus on the physically-motivated case of virialized pressure-bounded (i.e.~constant mean surface density) clouds (clumps).  Section \ref{sec:rdist} is the counterpart of Section \ref{sec:cc} as it models the radius distribution of CFRgs in relation to that of their parent clouds (clumps). Our conclusions are presented in Section \ref{sec:conclu}.

\section{Constant mean volume density for cluster-forming regions (CFRgs)}
\label{sec:evid}

The tidal field impact, namely, the ratio of the half-mass radius $r_h$ to the tidal radius $r_t$ of an embedded-cluster, quantifies  how deeply a cluster sits within its tidal radius and hence its likelihood of experiencing tidal overflow as it expands in response to gas-expulsion.  To satisfy the observed requirement of mass-independent cluster infant weight-loss, $r_h/r_t$ must be independent of the CFRg mass.  \citet{par10}  demonstrate that -- for given local SFE, gas expulsion time-scale and external tidal field --, this constrain is robustly satisfied for CFRgs with constant mean volume density.  This is because their half-mass radius $r_h$ and tidal radius $r_t$ scale alike with the embedded-cluster mass $m_{ecl}$, namely, $r_h \propto m_{ecl}^{1/3}$ and $r_t \propto m_{ecl}^{1/3}$.  For constant volume density cluster progenitors, the tidal field impact $r_h/r_t$ is thus mass-independent.  In contrast, constant surface density CFRgs lead to more massive clusters being more vulnerable to early destruction than their low-mass counterparts owing to a greater tidal field impact ($r_h \propto m_{ecl}^{1/2}$ and $r_h/r_t \propto m_{ecl}^{1/6}$), while the opposite is true for constant radius CFRgs ($r_h \propto m_{ecl}^{0}$ and $r_h/r_t \propto m_{ecl}^{-1/3}$).  Since observations suggest infant mortality/weight-loss to be mass-independent, the analysis performed by \citet{par10} lends strong support to the hypothesis that CFRgs have a constant mean volume density.    \\

In our Galaxy, observational evidence for CFRgs of constant mean volume density is provided by the tight association between postsigns of star formation ($IRAS$/$MSX$ sources, water masers, bipolar molecular outflows) and high density molecular gas, i.e. hydrogen molecule number densities $n_{H2} \simeq 10^{4-5}\,cm^{-3}$ \citep[or mean volume densities $\rho \simeq 700-7000\,M_{\sun}.pc^{-3}$;][]{eva08}.  \citet{aoy01} and \citet{yon05} note that star formation in $C^{18}O$ cores is often associated to the $H^{13}CO^{+}$ molecule, a tracer of molecular gas with $n_{H2} \simeq 10^5\,cm^{-3}$ (see bottom panel of Fig.~\ref{fig:isomth} and Section \ref{sec:cc}). 

That star formation requires a volume density threshold is also supported by studies of the molecular gas content and star formation activity of external galaxies.  \citet{gao04} obtain the $L_{IR}/L_{HCN}$ ratio of 65 infrared galaxies, where $L_{IR}$ is the galaxy-integrated infrared luminosity, and $L_{HCN}$ is the galaxy-integrated HCN J$=1-0$ line luminosity.  $L_{HCN}$ maps molecular gas with $n_{H2} \simeq 3 \times 10^4\,cm^{-3}$, while $L_{IR}$ traces the star formation rate (SFR).  From the near-constancy of $L_{IR}/L_{HCN}$, \citet{gao04} deduce that, on the average, galaxy-integrated SFRs scale linearly with their dense molecular gas content, from quiescent spirals to violent Ultra-Luminous Infra-Red Galaxies (ULIRGs).  In contrast, the ratio $L_{IR}/L_{CO}$, where the galaxy-integrated CO luminosity $L_{CO}$ traces molecular gas with $n_{H2} \simeq 300\,cm^{-3}$, is not on the average constant \citep[see figs~1 and 2 in][]{gao04}.  That is, $L_{HCN}$ traces the global SFR of galaxies better than $L_{CO}$.  

Using the same HCN J$=1-0$ molecular tracer as \citet{gao04}, \citet{wu05} map dense molecular clumps in Galactic GMCs.  They find that the one-to-one correlation between $L_{HCN}$ and $L_{IR}$ established by \citet{gao04} for entire galaxies also holds for individual dense molecular clumps.  \citet{gao04} and \citet{wu05} therefore argue that the most relevant parameter for the SFR is the amount of {\it dense} molecular gas, namely, gas with $n_{H2} \simeq 10^{4-5}\,cm^{-3}$.  For instance, the high SFR of ULIRGs stems from their large content of molecular gas with densities comparable to that of molecular clumps in Galactic GMCs (see also Section \ref{subsec:GMC}).  As pointed out by \citet{eva08}, these dense clumps provide the connection between star formation in the Milky Way and in other galaxies.  These conclusions are reminiscent of the earlier study of \citet{lad92}.   She find that, while the bulk of star formation in the Orion~B molecular cloud is associated with gas with $n_{H2} \simeq 10^4\,cm^{-3}$, the CO-traced gas is inert in terms of star formation.  \citet{lad10} achieve the same conclusion by comparing infrared extinction maps of local molecular clouds with their respective census of young stellar objects.  

A theoretical prediction of a number density threshold for star formation is made by \citet{elm07} who note that, when $n_{H2} \gtrsim 10^5\,cm^{-3}$, several microscopic effects enhance magnetic diffusion in the molecular gas, thereby significantly accelerating star formation (e.g. 
steeper density scaling for the electron fraction, 
modification of the coupling between dust grains and the magnetic field, 
sudden drop in the cosmic-ray ionization rate and hence in the ionization fraction) \citep[see][his section 3.6]{elm07}.  

That the bulk of star formation activity takes place in dense molecular gas regions characterised by a mean constant volume density is thus supported by both the analysis of the tidal field impact upon young clusters \citep{par10}, and by the tight association observed between star formation and dense molecular gas.  Given the uncertainties regarding the volume density threshold requested for star formation, calculations presented below are performed for two distinct cases: $n_{H2} \geq 10^4\,cm^{-3}$ \citep[e.g.][]{lad10} and $n_{H2} \geq 10^5\,cm^{-3}$ \citep[e.g.][]{elm07}.

\section{From molecular clumps to high-density cluster-forming regions}
\label{sec:cc}
Before going any further, a clarification of the terminology applied through this paper may be needed.  The following nomenclature has taken root in the community: the word `core' is now often restricted to the gaseous precursor of an individual star or of a small group of stars, while the term `clump' is designated for regions hosting cluster formation.  We will follow that terminology.  
The CFRg is the clump central region where active star formation takes place owing to a high enough volume/number density, i.e. $n_{H2} \geq 10^{4-5}\,cm^{-3}$. CFRg-related quantities are identified by the subscript `th', where `th' stands for (volume density) threshold.   
By virtue of the assumed spherical symmetry, a clump is assumed to contain one single forming-cluster.  This constitutes a major difference between the present model and the model by \citet{sha10} who consider the formation of many high-density regions in one single fractal cloud.  Note that the present terminology implies that the expressions `cluster-forming cores' used profusely in \citet{par08b}, \citet{par09} and \citet{par10} are now to be read `cluster-forming regions'.    
Although the subscript `clump' is used systematically in the equations below, these equations can be applied indifferently to any spherical volume of molecular gas containing a star-forming region in its central zone.  Should GMCs in galaxy starbursts and mergers be roughly spherical and forming each a massive star cluster in their centre, all equations developed in this paper can be applied to them.   \\

The mass function of molecular clumps and clouds mapped in $C^{18}O$, $^{13}CO$ or $^{12}CO$ emission line is well-described by a power-law $dN \propto m^{-\beta_0}dm$ with $\beta_0 \simeq 1.7$ \citep{kra98,won08}.  This is shallower than the `canonical' mass function $dN \propto m^{-\beta_\star}dm$ of young star clusters for which $\beta_\star \simeq 2$.  

\begin{figure}
\includegraphics[width=\linewidth]{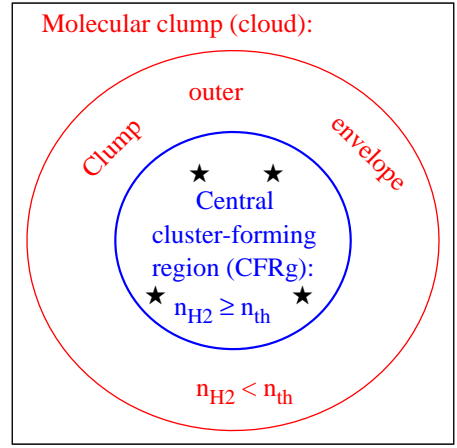}
\caption{Model of a molecular clump (cloud) hinged on through this paper.  The clump central region only, where the gas number density $n_{H2}$ achieves a threshold $n_{th}$, forms stars.  We refer to it as the `cluster-forming region' (CFRg).  Because the CFRg is defined based on a volume density threshold, its mass-radius relation is one of constant mean volume density, regardless of the host-clump mass-radius relation.  \label{fig:sketch} }
\end{figure}

To explain this difference in slope, our study rests on the clump (cloud) outer layers being inefficient at forming stars.  Let us consider the $C^{18}O$ clumps with masses $m_{clump}$ and radii $r_{clump}$ compiled in top and middle panels of Fig.~\ref{fig:isomth}.  This mass-radius diagram is based on the data of \citet[][the Orion~B molecular cloud, triangles]{aoy01}, \citet[][the GMC toward HII regions S35 and 37, squares]{sai99} and \citet[][the $\eta$~Carinae GMC, circles]{yon05} (see also section 3 in \citet{par10} for additional details).  By virtue of the molecular tracer used to map these clumps, their mean number densities sample the limited range $10^3\,cm^{-3} \lesssim n_{H2} \lesssim 10^4cm^{-3}$.  These density limits are shown as the dotted (black) lines in the middle panel of Fig.~\ref{fig:isomth}.  Let us consider a particular clump of mass $m_{clump}=1000\,M_{\sun}$ and radius $r_{clump}=1$\,pc.  Its mean number density $n_{H2} = 3.4 \times 10^3cm^{-3}$ suggests that it fails at forming stars if star formation actually requires a density threshold $n_{th} \simeq 10^{4-5}\,cm^{-3}$.  Yet, molecular clumps show density gradients (Section \ref{subsec:mod}) and the condition $n_{H2} \geq n_{th}$ may be met in the clump inner regions.  This in turn implies that the radius and mass of cluster gaseous progenitors are smaller than those of their host-clumps.  In what follows, $m_{th}$ and $r_{th}$ are the mass and radius of the CFRg where $n_{H2} \geq n_{th}$.  Conversely, clump outer layers are inefficient at forming stars owing to too low a volume density, i.e. $n_{H2} < n_{th}$.  This situation is illustrated in Fig.~\ref{fig:sketch}. 

The question we set to answer is: can the CFRg mass function $dN/dm_{th}$ and clump (cloud) mass function $dN/dm_{clump}$ differ?  As we shall see, it depends on the clump mass-radius relation and on the clump volume density profile.  

\subsection{Model for cluster-forming regions and their host-clumps}
\label{subsec:mod}
  
Let us characterise molecular clumps (clouds) with the following properties: \\
{\it (i)} Their volume density profile obeys a power-law of slope $-p$  
\begin{equation}
\rho _{clump}(s) = k_\rho \, s^{-p}\,,  
\label{eq:rho}
\end{equation}
with $s$ the distance from the clump centre and $k_\rho$ a normalizing factor.
The assumption of spherical symmetry is supported by e.g. the 1.2-mm continuum observations of \citet{bel06}
who find the mean and median ratios of the full widths at half maximum of their clumps along the $x$- and $y$-axes, $FWHM_x/FWHM_y$, to be 1.04 and 0.96, respectively.  Power-law density profiles for molecular clumps are put forward by various studies, e.g. \citet{hea93}, \citet{hat00}, \citet{beu02}, \citet{fon02} and \citet{mue02}.  Estimates for the density index $p$ are found mostly in the range $1.5 \lesssim p \lesssim 2.5$.  We insist that, in what follows, expressions  `constant volume density clumps' or `constant surface density clumps' do {\it not} imply that these clumps have a uniform volume or surface density.  Rather, it means a population of clumps all characterized by the same {\it mean} surface or volume density. \\

{\it (ii)} The mass-radius relation of molecular clumps is quantified by its slope $\delta$ and normalization $\chi$:
\begin{equation}
r_{clump}[pc] = \chi (m_{clump}[M_{\sun}])^{\delta}\,.
\label{eq:rcmc}
\end{equation}
The combination of Eqs.~\ref{eq:rho} and \ref{eq:rcmc} leads to the CFRg mass $m_{th}$, i.e. the mass of the clump region where $\rho_{clump}(s) \geq \rho_{th}$.  Equation \ref{eq:mth} provides $m_{th}$ as a function of the clump mass $m_{clump}$, radius $r_{clump}$ and density index $p$:
\begin{equation}
\label{eq:mth}
m_{th} = \left( \frac{3-p}{4 \pi \rho _{th}} \right)^{(3-p)/p} m_{clump}^{3/p} \, r_{clump}^{-3(3-p)/p}\,.
\end{equation}
It is valid for density indices $0 < p <3$.
Similarly, the radius $r_{th}$ of the spherical CFRg is:
\begin{equation}
\label{eq:rth}
r_{th} = \left( \frac{3-p}{4 \pi \rho _{th}} \right)^{1/p} m_{clump}^{1/p} \, r_{clump}^{-(3-p)/p}\,.
\end{equation}
It immediately follows that the CFRg mean density $<\rho _{th}>$ is constant.  It depends solely on the density index $p$ and  density threshold $\rho _{th}$:
\begin{equation}
<\rho _{th}> = \frac{3 m_{th}}{4\pi r_{th}^3} = \frac{3}{3-p} \rho _{th}\,.
\label{eq:av_rhoth}
\end{equation}
In other words, the mass-radius relation of CFRgs obeys $r_{th} \propto m_{th}^{1/3}$.  This is a key-point for our forthcoming discussion about how clump and CFRg mass functions differ from each other.  Equation \ref{eq:av_rhoth} shows that for a truncated isothermal sphere ($p=2$), the CFRg mean volume density is three times higher than the threshold $\rho _{th}$.  For shallower density indices, the density contrast is weaker, e.g. $<\rho _{th}> / \rho _{th}=2$ when $p=1.5$.
 
\begin{figure}
\includegraphics[width=\linewidth]{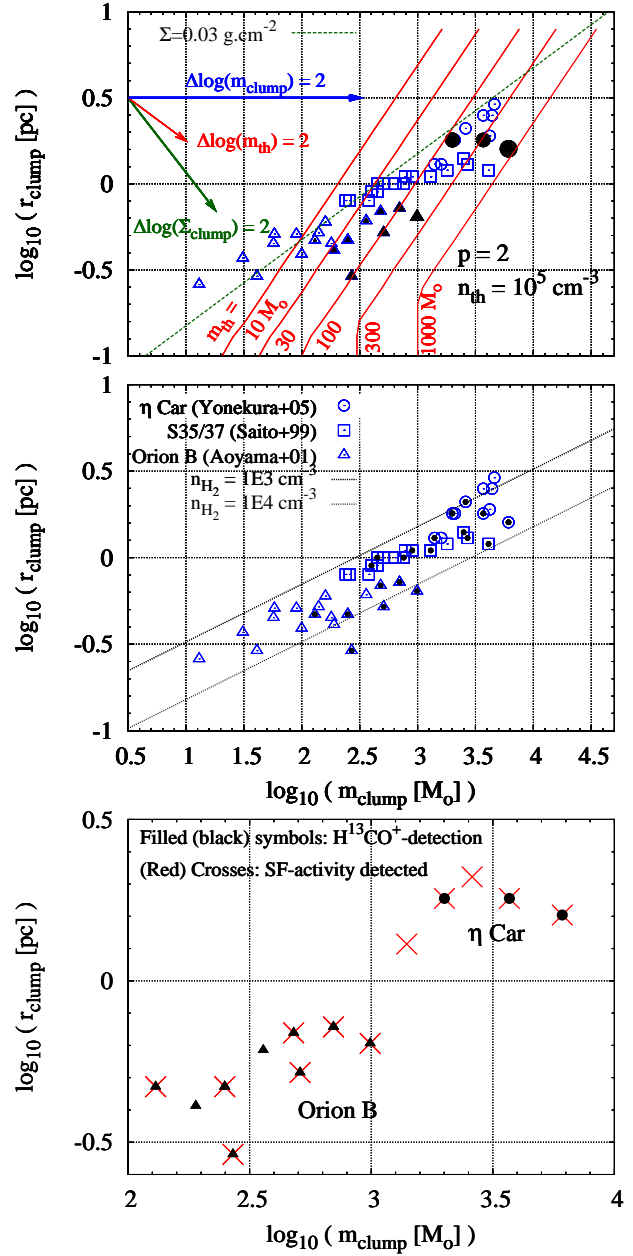}
\caption{Radius vs. mass of molecular $C^{18}O$ clumps (triangles, squares, circles; see text for details).  {\it Top panel:} Filled symbols indicate an $H^{13}CO^+$ detection, with the filled symbol size scaling with the mass of $H^{13}CO^+$-detected gas. No $H^{13}CO^+$-data are available for the open squares.  Solid (red) lines are iso-$m_{th}$ lines, where $m_{th}$ is the predicted mass of the central CFRg, assuming a density index $p=2$ and $n_{th} = 10^5\,cm^{-3}$.  From left to right: $m_{th}=10, 30, 100, 300, 1000\,M_{\sun}$.  Arrows indicate 2-order-of-magnitude increases of the (from top to bottom) clump mass, CFRg mass and clump surface density.  {\it Middle panel:} Same $C^{18}O$ clump mass-radius diagram as above.  Black dots indicate clumps with detected star formation activity.  Also indicated is the range of their mean number densities, limited to $\simeq 10^3$-$10^4\,cm^{-3}$  by virtue of the molecular tracer ($C^{18}O$) used to map them.  {\it Bottom panel:} 
Comparison between detections in $H^{13}CO^+$-emission (filled symbols from the top panel) and detections of star formation activity (the crosses equate with the middle panel filled circles) for the clumps in Orion B (triangles) and $\eta$ Car (circles).  The data of S35/37 (open squares in top and  middle panels) are not shown since those have not been mapped in $H^{13}CO^+$.  Note that the $x$- and $y$-spans differ from those in top and middle panels.  \label{fig:isomth} }
\end{figure}

To quantify CFRg masses, an estimate of $\rho_{th}$ (or $n_{th}$) is needed.  
In the middle panel of Fig.~\ref{fig:isomth}, open symbols marked with a black filled circle indicate $C^{18}O$ clumps with detected star formation activity ($IRAS$ or $MSX$ source, or bipolar molecular outflow).  Data come from tables 1-3 in \citet{aoy01}, tables 2 and 4 in \citet{sai99}, and table 3 in \citet{yon05}.  These C$^{18}O$ clumps were also observed by \citet{aoy01} and \citet{yon05} in the H$^{13}$CO$^+$ $J =1-0$ emission line so as to detect  $n_{H_2} \simeq 10^5\,cm^{-3}$ gas.  Filled symbols in top panel of Fig.~\ref{fig:isomth} highlight $C^{18}O$ clumps detected in $H^{13}CO^+$-emission.  Symbol size is proportional to the mass of $H_2$ detected in $H^{13}CO^+$-emission.  No $H^{13}CO^+$ data is provided by \citet{sai99}.  
The bottom panel of Fig.~\ref{fig:isomth} zooms in on the $H^{13}CO^+$-data and the star formation activity detections for Orion~B and $\eta$ Car [the S35/37 data of \citet{sai99} are ignored as they lack an $H^{13}CO^+$-mapping].  The comparison between both types of data highlights the tight correlation between H$^{13}$CO$^+$-detected gas and star formation activity, a point already made by \citet{aoy01} and \citet{yon05}: 10 in 12 $H^{13}CO^+$-detections also show signs of star formation activity, while 10 in 12 $C^{18}O$ clumps with detected star formation activity also host an $H^{13}CO^+$-detected region.  Similarly, \citet{hig10} detect $H^{13}CO^+$-emission in the $C^{18}O$ clumps associated to embedded clusters studied by \citet{hig09} (see also Section \ref{subsec:clumps}).
This suggests that star formation requires number densities of order $n_{H_2}\simeq10^5\,cm^{-3}$ (or $\rho \simeq 7000\,M_{\sun}.pc^{-3}$).  We therefore adopt $10^5\,cm^{-3}$ as the fiducial $n_{th}$ \citep[see also][]{elm07}.  We will also present results for $n_{th} = 10^4\,cm^{-3}$ \citep{lad10}.  

Note that the presence of $C^{18}O$ clumps with mean densities $\gtrsim 10^3\,cm^{-3}$ and hosting star formation activity (middle panel of Fig.~\ref{fig:isomth}) does not imply that star formation can take place at so low number densities.  Molecular clumps are characterized by density gradients and star formation is very likely confined to the clump deeper regions (Fig.~\ref{fig:sketch}) whose higher volume densities are revealed in H$^{13}$CO$^+$.       \\

\begin{figure}
\includegraphics[width=\linewidth]{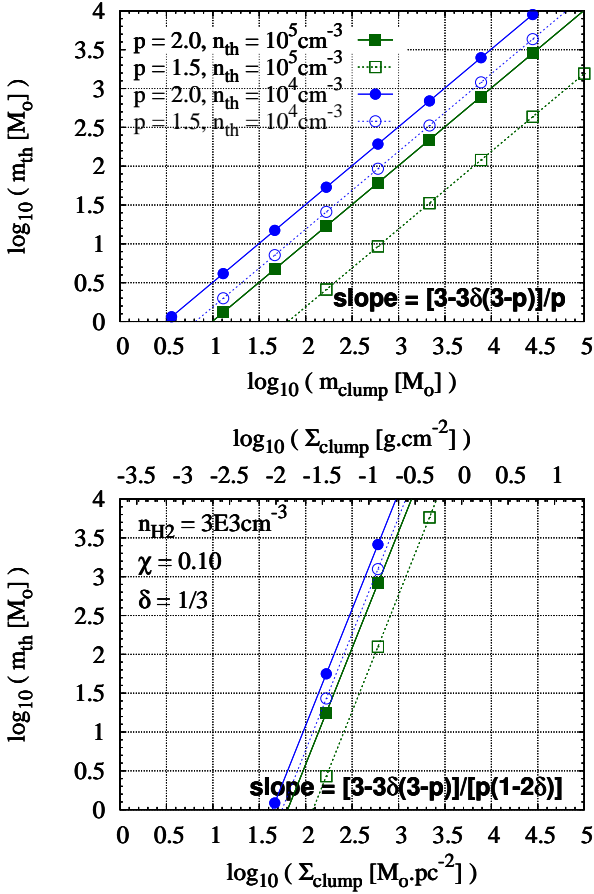}
\caption{{\it Top panel:} Relation between the mass of the high-density cluster-forming region $m_{th}$ and the total clump mass $m_{clump}$.  {\it Bottom panel:} The mass of the high-density cluster-forming region $m_{th}$ plotted against the average clump surface density $\Sigma_{clump}$.  Model parameters $(p, n_{th}, \chi, \delta)$ are quoted in the figure, and the slope of each relation is quoted in its respective panel.   Note that the dependence on $\log(\Sigma_{clump})$ is steeper by a factor $(1-2\delta)^{-1}$ than the dependence on $\log(m_{clump})$.  \label{fig:sig_dep} }
\end{figure}

Solid (red) lines in the top panel of Fig.~\ref{fig:isomth} are iso-$m_{th}$ lines in the $m_{clump}$ vs.~$r_{clump}$ space when $p=2$  and $n_{th}=10^5\,cm^{-3}$ (Eq.~\ref{eq:mth}).  In our model, the total stellar mass inside a clump scales with the mass $m_{th}$ of its central CFRg rather than with the clump total mass $m_{clump}$.  Given the CFRg (local) SFE, the stellar mass $m_{ecl}$ of the embedded-cluster is $SFE \times m_{th}$.  The top panel of Fig.~\ref{fig:isomth} also displays vectors along which $m_{th}$, $m_{clump}$ and $\Sigma _{clump}$ increase by 2 orders of magnitude, with $\Sigma _{clump} = m_{clump}/(\pi r_{clump}^{\,2})$ the average clump surface density.  These vectors illustrate that the CFRg mass $m_{th}$ depends more sensitively on the average clump surface density $\Sigma _{clump}$ than on $m_{clump}$, an effect depicted in Fig.~\ref{fig:sig_dep}.  Its top and bottom panels show the mass of the high-density CFRg $m_{th}$ as a function of the total clump mass $m_{clump}$ and of the average clump surface density $\Sigma_{clump}$, respectively.  These relations are obtained by combining Eqs.~\ref{eq:rcmc} and \ref{eq:mth}, and their logarithmic slopes are quoted in their respective panels.  Four models combining $n_{th}=10^4\,cm^{-3}$ or $n_{th}=10^5\,cm^{-3}$, with $p=1.5$ or $p=2.0$ are presented.  In all cases, the clumps are assumed to have a constant mean volume density, namely, $n_{H2}=3 \times 10^3\,cm^{-3}$, typical of $C^{18}O$ clumps.  As suggested by the vectors in the top panel of Fig.~\ref{fig:isomth}, the slope of the $\log(m_{th})$ vs $\log(\Sigma_{clump})$ relation is steeper than its counterpart $\log(m_{th})$ vs $\log(m_{clump})$, by a factor $(1-2\delta)^{-1}$.  As an example, for clumps with a given mean volume density ($\delta = 1/3$), the ratio between both slopes is a factor of 3.  The embedded-cluster stellar mass $SFE \times m_{th}$, thus also its luminosity, therefore depend more sensitively on $\Sigma _{clump}$ than on $m_{clump}$.  This explains straightforwardly why \citet{aoy01} find the IRAS luminosity of their $C^{18}O$ clumps `to be more strongly dependent on the average column density $\Sigma _{clump}$ than on the total mass of the clump' $m_{clump}$.  In a forthcoming paper, we will model the infrared luminosity of molecular clumps as a way of probing their forming stellar content and compare model outputs with existing data-sets.

The lowest-mass Orion~B clumps (with $m_{clump} \lesssim 150\,M_{\sun}$) show neither sign of star formation nor $H^{13}CO^+$-detection.  The top panel of Fig.~\ref{fig:isomth} shows that if $p=2$, these clumps contain $m_{th} \lesssim 20\,M_{\sun}$ of high-density gas.  So low a mass may result in neither (detected) star formation, nor $H^{13}CO^+$ detection.

\subsection{From the clump mass function to the CFRg mass function}
\label{subsec:mf}
Having defined the properties of CFRgs and of their host-clumps, we are now ready to relate the CFRg mass function to the clump mass function.  Our model explicitly assumes that the high-density CFRgs -- the genuine sites of cluster formation -- have a constant mean volume density (Eq.~\ref{eq:av_rhoth}) and, thus, that their mass-radius relation scales as $r_{th} \propto m_{th}^{1/3}$.  In this section, we show that if the mass-radius relation of their host-clumps has a different slope (i.e.~$\delta \ne 1/3$), the slope of the CFRg mass function differs from that of the clump mass function too (i.e. $\beta \neq \beta_0$). \\

\begin{figure}
\includegraphics[width=\linewidth]{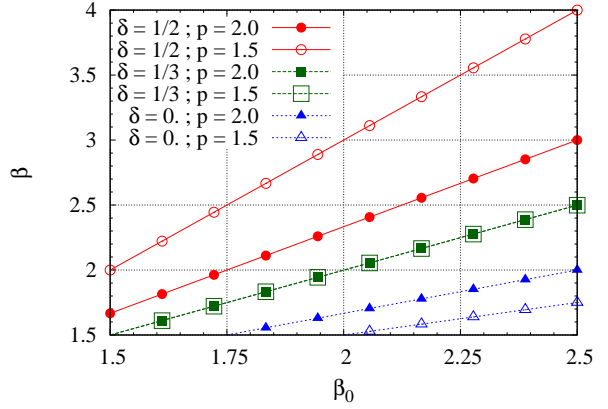}
\caption{Index $\beta$ of the CFRg mass function in dependence of the index $\beta _0$ of the clump mass function for three clump mass-radius relations -- constant mean surface density ($\delta =1/2$), constant mean volume density ($\delta =1/3$) and constant radius ($\delta =0$) -- and two density indices ($p=2$ and $p=1.5$).  Note that a constant local (i.e. CFRg) SFE implies $\beta$ to also be the index $\beta_\star$ of the embedded-cluster mass function. \label{fig:beta} }
\end{figure}

As a first step, this can be understood by obtaining the clump mass fraction occupied by the dense central CFRg.
Combining Eqs.~\ref{eq:rcmc} and \ref{eq:mth}, we obtain this mass ratio
\begin{equation}
\label{eq:mth/mc}
\frac{m_{th}}{m_{clump}} = \left( \frac{3-p}{4 \pi \rho _{th} \chi^3} \right)^{(3-p)/p} m_{clump}^{[(3-p)(1-3\delta)]/p}
\end{equation}
as a function of normalization $\chi$ and slope $\delta$ of the clump mass-radius relation (Eq.~\ref{eq:rcmc}), of the slope $p$ of the clump density profile (Eq.~\ref{eq:rho}), of the clump mass $m_{clump}$, and of the volume density $\rho_{th}$ at the edge of the CFRg.  With the normalization $\chi$ of Eq.~\ref{eq:rcmc}, $\rho_{th}$ and $m_{clump}$ are in units of $M_{\sun}.pc^{-3}$ and $M_{\sun}$, respectively. \\
For constant volume density clumps ($\delta=1/3$), this mass fraction is independent of $m_{clump}$
\begin{equation}
\label{eq:mth/mcrho}
\frac{m_{th}}{m_{clump}} = \left( \frac{3-p}{4 \pi \rho _{th} \chi^3} \right)^{(3-p)/p}
\end{equation}
and the clump and CFRg mass function slopes are thus alike.
For constant surface density clumps ($\delta=1/2$), this mass fraction 
\begin{equation}
\label{eq:mth/mcSigma}
\frac{m_{th}}{m_{clump}} = \left( \frac{3-p}{4 \pi \rho _{th} \chi^3} \right)^{(3-p)/p} m_{clump}^{-(3-p)/(2p)}
\end{equation}
is a decreasing function of $m_{clump}$ (since $0<p<3$).  Therefore, we predict a CFRg mass function {\it steeper} than the clump mass function. 
For constant radius clumps ($\delta=0$), the CFRg mass fraction obeys  
\begin{equation}
\label{eq:mth/mcr}
\frac{m_{th}}{m_{clump}} = \left( \frac{3-p}{4 \pi \rho _{th} \chi^3} \right)^{(3-p)/p} m_{clump}^{(3-p)/p}\,.
\end{equation}
It is an increasing function of $m_{clump}$, rendering the CFRg mass function {\it shallower} than the clump mass function.  These effects are expected since, for constant surface density clumps, the mean volume density decreases with increasing mass, while the opposite is true for constant radius clumps.   \\ 

Let us now quantify these effects in detail and let us consider a population of clumps whose mass distribution is a power-law of slope $-\beta _0$:
\begin{equation}
dN = k_{clump} m_{clump}^{-\beta _0} dm_{clump}\;.
\label{eq:coMF}
\end{equation}
To derive the CFRg mass function 
\begin{equation}
dN = k_{th} m_{th}^{-\beta} dm_{th}\,,
\end{equation}
we need to relate the CFRg mass $m_{th}$ to the clump mass $m_{clump}$.  Equation~\ref{eq:mth/mc} straightforwardly leads to: 
\begin{equation}
\label{eq:mthmc}
m_{th} = \left( \frac{3-p}{4 \pi \rho _{th} \chi^3} \right)^{(3-p)/p} m_{clump}^{[3-3 \delta (3-p)]/p}\,.
\end{equation}

\begin{figure}
\includegraphics[width=\linewidth]{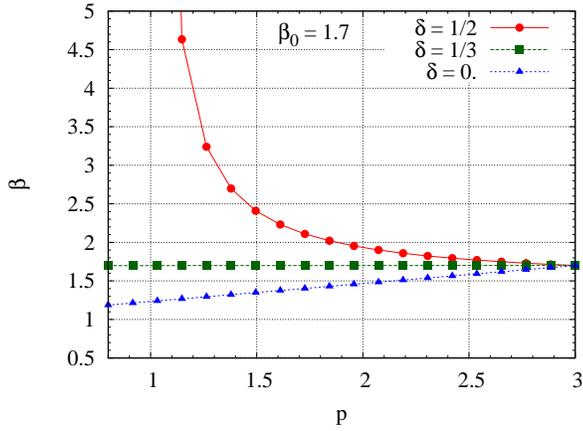}
\caption{How the index $\beta$ of the CFRg mass function differs from an assumed clump mass function index $\beta _0=1.7$ as a function of the density index $p$.  Clump mass-radius relations are as in Fig.~\ref{fig:beta} \label{fig:betap} }
\end{figure}

The combination of Eqs.~\ref{eq:coMF} and \ref{eq:mthmc} leads to the power-law mass function of CFRgs: 
\begin{eqnarray}
dN  = k_{clump} \left( \frac{4 \pi \rho _{th} \chi^3}{3-p} \right)^{ (3-p)(1-\beta _0)/[3-3\delta (3-p)] } \nonumber \\
\frac{p}{3-3\delta (3-p)} \, m_{th}^{-\beta}  dm_{th}\,.
\label{eq:MFth}
\end{eqnarray}
with $\beta$ obeying:
\begin{equation}
\label{eq:slopes}
\beta = \frac{p\beta _0 - (p-3)(1-3\delta )}{3-3\delta (3-p)}
\end{equation} 
Figure~\ref{fig:beta} shows $\beta$ in dependence of $\beta _0$ for 6 distinct cases: clumps with constant surface density ($\delta =1/2$), constant volume density ($\delta =1/3$) and constant radius ($\delta =0$), combined to two density indices: $p=2$ (isothermal spheres) and $p=1.5$.  As we saw above, $\delta =1/3$ leads to $\beta = \beta _0$, while constant clump surface density (radius) increases (decreases) $\beta$ compared to $\beta _0$.  Shallower clump density profiles (open symbols in Fig.~\ref{fig:beta}) result in a greater contrast between the clump and CFRg mass function slopes, that is, $|\beta - \beta _0|$ is greater for smaller density index $p$.
This effect is further quantified in Fig.~\ref{fig:betap} which depicts $\beta$ as a function of $p$ for a given spectral index $\beta _0=1.7$ of the clump mass function.  $\delta =1/3$ is conducive to $\beta = \beta _0 = 1.7$.  When $p \rightarrow 3$, the difference between the clump and CFRg mass function slopes vanishes, irrespective of the clump mass-radius relation.  In contrast, the smaller the density index $p$, the steeper (shallower) the CFRg mass function when $\delta =1/2$ ($\delta =0$).    \\ 

\begin{figure}
\includegraphics[width=\linewidth]{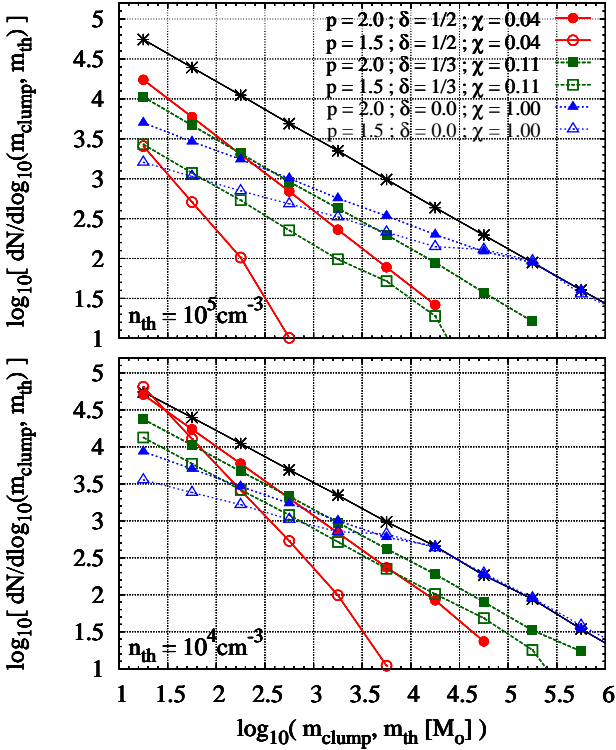}
\caption{Comparison between the clump mass spectrum (solid black lines with asterisks) and CFRg mass spectra.  The clump spectral index is $\beta _0=1.7$.  Symbol- and colour-codings are identical to Fig.~\ref{fig:beta}.
$\chi$ is the normalization of the clump mass-radius relation (Eq.~\ref{eq:rcmc}) \label{fig:MF} }
\end{figure}

Figure~\ref{fig:MF} presents the outcome of Monte-Carlo simulations performed to compare CFRg mass functions to their `parent' clump mass function.  In each panel, the latter is depicted as the upper black line with asterisks.  The same six combinations of $\delta$ and $p$ as previously are considered and symbol/colour-coding is identical  to Fig.~\ref{fig:beta}.  Adopted normalizations $\chi$ for the clump mass-radius relation (Eq.~\ref{eq:rcmc}) correspond to the best-fits of the $C^{18}O$ data in Fig.~\ref{fig:isomth} with slope $\delta$ imposed.  Those were obtained by \citet[][their table 1]{par10}.  We remind those clump mass-radius relations below for the sake of clarity.  Fitting a constant surface density relation onto the $C^{18}O$ data leads to $\chi=0.04$:
\begin{equation}
r_{clump}=0.04 m_{clump}^{1/2}\,,
\end{equation}  
equivalent to $\Sigma _{clump}=0.05g.cm^{-2}$.  
A constant volume density fit results in $\chi=0.11$:
\begin{equation}
r_{clump}=0.11 m_{clump}^{1/3}\,,
\end{equation}  
or $n_{H_2, clump}=3.10^3\,cm^{-3}$.

For constant clump radius, we adopt $r_{clump}=1pc$.

As discussed in Section \ref{subsec:mod}, two distinct volume density thresholds are considered: $n_{th} = 10^5\,cm^{-3} \equiv \rho_{th} = 7000\,M_{\sun}.pc^{-3}$ (top panel of Fig.~\ref{fig:MF}), and $n_{th} = 10^4\,cm^{-3} \equiv \rho_{th} = 700\,M_{\sun}.pc^{-3}$ (bottom panel of Fig.~\ref{fig:MF}).

Not only do shallower clump density profiles lead to greater $|\beta - \beta_0|$, they are also conducive to lower normalizations of the CFRg mass function compared to the clump mass function.  This effect stems from a smaller clump mass fraction achieving the volume density threshold for lower $p$.   
The horizontal shift between the clump and CFRg mass functions also depends on $\rho_{th}$ and $\chi$ (see Eq.~\ref{eq:mth/mc}).  The closer to $\rho_{th}$ the clump mean volume density is, the greater the clump mass fraction contained within the CFRg and the smaller the horizontal shift between the clump and CFRg mass functions.  When the whole clump is at a density of at least $\rho_{th}$, clump and CFRg mass functions coincide (e.g. when $r_{clump}=1\,pc$, $n_{th}=10^4cm^{-3}$ and $log_{10}(m_{clump})\gtrsim 4$, see lines with triangles in bottom panel of Fig.~\ref{fig:MF}).     

That CFRgs hosted by constant surface density clumps ($\delta=1/2$) have a steeper mass spectrum than their parent clumps is a highly interesting result since the same is observed for gas-embedded clusters and young gas-free clusters ($\beta_\star=2$) as compared to GMCs and their dense gas clumps ($\beta_0=1.7$).  We discuss this issue in Section~\ref{sec:implic}.   

\subsection{`Global' and `local' SFEs}
\label{subsec:sfe}
That only a limited region of a molecular clump may form a star cluster means that the SFE relevant to model cluster violent relaxation must be defined properly.  The dynamical response of a cluster to the expulsion of its residual star-forming gas is governed in part by the mass fraction of gas turned into stars {\it within the volume of gas forming stars} and {\it at the onset of gas expulsion}\footnote{In this contribution, we assume that stars and gas in the forming-cluster are in virial equilibrium at gas expulsion onset.  This implies that the gas mass fraction turned into stars within the CFRg equates with the effective SFE \citep[eSFE,][]{goo09}.  That is, the CFRg (or local) SFE plays a key-role in the early evolution of star clusters.  For detailed discussions of this assumption, see \citet{kro08} and \citet{goo09}.}.  We refer to this SFE as the `local' SFE $\epsilon _{local}$, namely, the CFRg mass fraction eventually turned into stars.  Conversely, an SFE averaged over the whole molecular clump -- hereafter `global' SFE $\epsilon_{global}$ -- constitutes a lower limit only to the local SFE.  

How the local SFE ($\epsilon _{local} = m_{ecl}/m_{th}$), global SFE ($\epsilon _{global} = m_{ecl}/m_{clump}$), and CFRg mass fraction $m_{th}/m_{clump}$ connect the embedded-cluster mass function $dN/dm_{ecl}$, the CFRg mass function $dN/dm_{th}$ and the clump mass function $dN/dm_{clump}$ is summarised in Fig.~\ref{fig:skMF}.

Using Eq.~\ref{eq:mth/mc}, it is straightforward to relate these two SFEs:
\begin{eqnarray}  
\epsilon _{global} = \frac{m_{ecl}}{m_{clump}} = \frac{m_{ecl}}{m_{th}} \frac{m_{th}}{m_{clump}} =  \epsilon _{local} \frac{m_{th}}{m_{clump}}  \nonumber \\
= \epsilon _{local} \left(\frac{3-p}{4\pi \rho _{th} \chi ^3}\right)^{(3-p)/p} m_{clump}^{[(3-p)(1-3\delta)]/p}\,. 
\label{eq:sfe}
\end{eqnarray}
Given a density threshold $\rho_{th}$, a lower normalisation $\chi$ is conducive to higher-density clumps, larger clump mass fractions with $\rho \geq \rho _{th}$ and, therefore, higher global SFEs.   

\begin{figure}
\includegraphics[width=\linewidth]{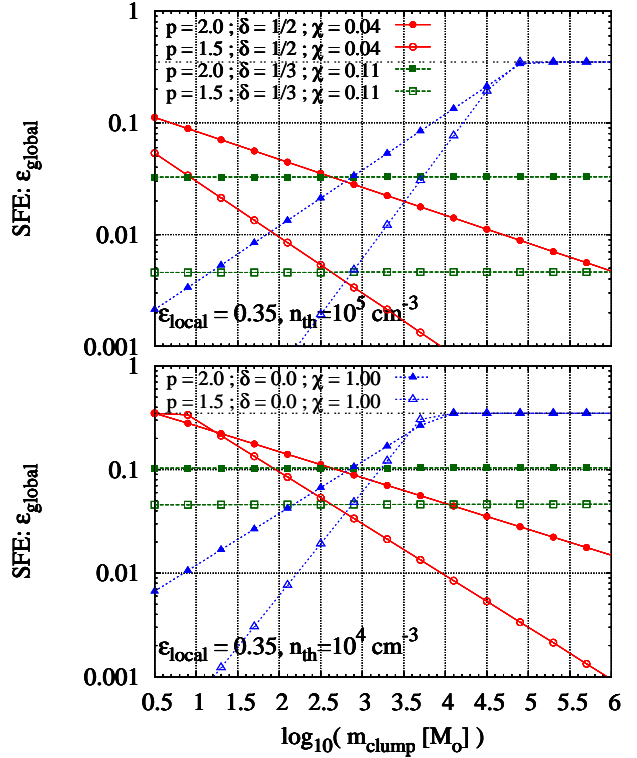}
\caption{Global SFE (i.e. SFE averaged over the whole clump) as a function of clump mass in the case of a spherical clump hosting a forming-cluster in its central region (see Fig.~\ref{fig:sketch}).  Symbol- and colour-codings are identical to Figs.~\ref{fig:beta} and \ref{fig:MF}.  The local SFE $\epsilon_{local}$ (i.e.~the SFE of the CFRg) is 0.35 (indicated by the horizontal dotted black line).  Cluster violent relaxation is determined by the local SFE, not the global one.  \label{fig:sfe} }
\end{figure}

Figure~\ref{fig:sfe} shows Eq.~\ref{eq:sfe} for the same ($p$, $\delta$, $\chi$, $n_{th}$) sets as in Fig.~\ref{fig:MF}, with identical colour- and symbol-codings.  The adopted local SFE is $\epsilon_{local}=0.35$.  It is shown as a horizontal dotted (black) line in both panels.  For a weak external tidal field impact, this SFE ensures that the cluster survives violent relaxation, even if the gas expulsion time-scale is much  shorter than a CFRg crossing-time \citep[i.e.~explosive gas expulsion; see fig.~1 in ][]{par07}.  Yet, Fig.~\ref{fig:sfe} illustrates that the global SFE measured over an entire $C^{18}O$ clump can be significantly smaller than $\epsilon_{local}$ and misleadingly suggests that the embbeded-cluster is not to survive violent relaxation.  Therefore, small SFEs for $C^{18}O$ clumps reported in the literature \citep[e.g.][their table 3]{hig09} do not necessarily imply that embedded-clusters get disrupted after gas expulsion.

Besides, one should also keep in mind that an observed SFE may be low because the CFRg is still in the process of building up its stellar content.  This trend is observed by \citet{hig09} who mapped in $C^{18}O$-emission 14 molecular clumps associated to embedded-clusters. They define a sequence A-B-C of $C^{18}O$ clump morphology (their table 3).  In Type-A clumps, the cluster is associated with the peak of $C^{18}O$ emission.  In contrast, clusters hosted by Type-C clumps are located at a cavity-like $C^{18}O$-emission hole, which demonstrates that gas dispersal has started in Type-C clumps.  \citet{hig09} therefore conclude that the morphology sequence A-B-C equates with an evolutionary sequence, with Type-A and Type-C corresponding to the least and most evolved clumps, respectively.  In further support of their scenario, they find a trend for the global SFE to increase along the sequence A-B-C.  The key-point to keep in mind here is that cluster violent relaxation depends on the local SFE -- i.e.~within the CFRg -- {\it at the onset of gas expulsion}.  
 
Figure~\ref{fig:sfe} illustrates that, under the assumption of a constant $\epsilon_{local}$, $\epsilon_{global}$ is clump-mass dependent when $\delta \neq 1/3$.  As this is the local SFE which rules cluster violent relaxation, the potential dependence of the global SFE on the clump mass is in itself {\it not} conducive to mass-dependent effects during violent relaxation.  If the local SFE, gas-expulsion time-scale in units of a CFRg crossing-time \citep{par08b} and tidal field impact \citep{par10} are CFRg-mass-independent, the slope of the cluster mass function through violent relaxation does not change, regardless of whether the global SFE is clump-mass-dependent or not. \\

The cases illustrated in Fig.~\ref{fig:sfe} assume that one clump hosts one CFRg, as depicted in Fig.~\ref{fig:sketch}.  The structure of GMCs in spiral galaxies, however, is more complex as one GMC hosts several clumps, possibly strung out on a filament, each hosting a star-forming or cluster-forming region.  For instance, let us consider the case of a $10^5M_{\odot}$ GMC hosting 10 dense clumps characterized by a constant mean surface density $\Sigma _{clump}=0.1\,g.cm^{-2}$ (or $\chi=0.025$) and density index $p=2$.  That surface density is characteristic of $C^{18}O$ clumps showing signs of star-formation activity (see top panel of Fig.~\ref{fig:isomth}).  The random sampling of a clump mass function with slope $-1.7$ and mass lower limit $100\,M_{\odot}$ shows that those clumps totalize on the average $\simeq 4 \times 10^4\,M_{\odot}$ of molecular gas.  Assuming a star formation density threshold of $n_{th} \simeq 3\times 10^4\,cm^{-3}$ (i.e. at the logarithmic midpoint between the two values tested through this paper), the total mass in dense star-forming gas represents only one tenth of the total clump mass, that is, $\simeq 4 \times 10^3\,M_{\odot}$.  Over the scale of the whole GMC, the filling factor for the star-forming dense gas is thus $\simeq 4 \times 10^3\,M_{\odot}/10^5M_{\odot} \simeq 0.04$.  Assuming $\epsilon _{loc}=0.35$, the GMC global SFE is $0.04\epsilon _{loc} \simeq 0.01$, a value typical for Galactic GMCs \citep{due82}.    

In galaxy starbursts and mergers, GMCs get compressed by the high pressures pervading such violent environments.  This results in high volume densities through most of the GMC volume.  In Section \ref{subsec:GMC}, we will speculate that such a GMC is roughly spherical with a smooth density profile and the birth site of one single massive cluster in its central regions.  That is, the relations established in this section remain valid and are transposed to the case of a GMC.

\section{Model Consequences}
\label{sec:implic}

Figures \ref{fig:beta}, \ref{fig:betap} and \ref{fig:MF} demonstrate that the mass distribution from molecular clumps (clouds) to CFRgs steepens ($\beta > \beta_0$) {\it if clumps (clouds) have a constant mean surface density}.  We remind here that, in our model, CFRgs have a constant mean {\it volume} density by virtue of the assumed volume density threshold for star formation (Eq.~\ref{eq:av_rhoth}).  The mean density index found by \citet{mue02} for the star-forming regions they map in dust-continuum emission is $p=1.8$.  Combined to $\beta _0 = 1.7$ for the clump mass function, this is conducive to $\beta \simeq 2$ for the CFRg mass function (see Fig.~\ref{fig:betap}).  Other realistic values of $p$ (see references in Section \ref{subsec:mod}) lead to steepenings ranging from $\beta \simeq 1.8$ when $p=2.5$ (an effect probably undetectable amidst data noise) to $\beta \simeq 2.4$ when $p=1.5$.  These $\beta$ values bracket the `canonical' mass function slope of young star clusters ($\beta_{\star} \simeq 2$), provided that the `local' SFE is mass-independent (i.e. $\beta = \beta_\star$).  The key-point to investigate now is whether molecular clumps and clouds hosting CFRgs have a constant mean surface density.  We consider two distinct cases: GMCs and their dense clumps.  In quiescent spirals such as our Galaxy, open clusters are observed to form within dense clumps in GMCs.  In galaxy starbursts and mergers, compressed GMCs are likely the individual birth-sites of massive star clusters forming profusely in these violent star-forming environments.       

\subsection{GMCs as cluster-forming sites}
\label{subsec:GMC}

\begin{figure}
\includegraphics[width=\linewidth]{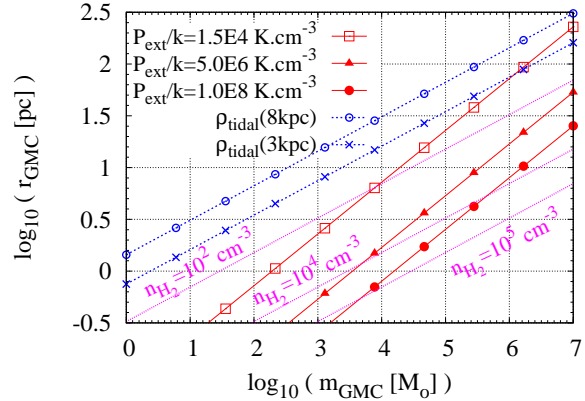}
\caption{Mass-radius relations of constant mean surface density clouds (solid (red) lines) for 3 distinct external pressures: $P_{ext}/k=1.5 \times 10^4 K.cm^{-3}$ (Galactic disc), $P_{ext}/k = 5 \times 10^6 K.cm^{-3}$ (Galactic bulge), and $P_{ext}/k=10^8 K.cm^{-3}$ (galaxy mergers).  The volume density limits imposed by the Galactic tidal field at galactocentric distances of 8\,kpc and 3\,kpc are shown as the (blue) dashed lines.  The two rightward dotted (pink) lines denote constant volume densities $n_{H2}=10^4cm^{-3}$ and $n_{H2}=10^5cm^{-3}$, typical of star-forming regions. \label{fig:mrGMC} }
\end{figure}

Pressure-bounded clouds in virial equilibrium have a constant mean surface density, provided that the external pressure $P_{ext}$ is about constant for all clouds \citep[see e.g.][]{hp94}:
\begin{equation}
r_{cloud} \propto m_{cloud}^{1/2} P_{ext}^{-1/4}\,.
\end{equation}
GMCs in our Galaxy occupy a narrow range in mass surface density $\Sigma_{GMC}$ with $10 \lesssim \Sigma_{GMC} \lesssim 100\,M_{\sun}.pc^{-2}$ \citep[fig.~8 in][]{bli06}.  More recently, \citet{hey09} found that, on the   average, $\Sigma_{GMC} \simeq 40\,M_{\sun}.pc^{-2}$ \citep[see also][]{lom10}.  Given the pressure characterising the Galactic disc, this is about the surface density expected for virialized gas clouds in pressure equilibrium with their environment.  The cloud external pressure $P_{ext}$ and cloud surface density $\Sigma_{GMC}$ are related through \citep{hp94}:
\begin{equation}
\Sigma_{GMC}=0.5 \left( \frac{P_{ext}}{k} \right)^{1/2} M_{\sun}.pc^{-2}\,.
\end{equation}   
The pressure in the Galactic disc is $P_{ext}/k \simeq 1.5 \times 10^4 K.cm^{-3}$ \citep{bli04}, leading to $\Sigma_{GMC}=60M_{\sun}.pc^{-2}$ (or $\Sigma_{GMC}=1.3 \times 10^{-2} g.cm^{-2}$).  This is in excellent agreement with the estimate of \citet{hey09}.
A surface density $\Sigma_{GMC} \simeq 60\,M_{\sun}.pc^{-2}$ equates with the mass-radius relation $(r_{GMC}/1pc) = 0.07\,(m_{GMC}/1M_{\sun})^{1/2}$, shown as the solid (red) line with open squares in Fig.~\ref{fig:mrGMC}.  

The mean number density of Galactic GMCs spans the range $10^1\,cm^{-3} \lesssim n_{H2,GMC} \lesssim 10^3\,cm^{-3}$ (see Fig.~\ref{fig:mrGMC}).  This is far too low for them to experience overall efficient star formation.  The situation in galaxy starbursts and mergers is drastically different, however, because external pressures there are several orders of magnitude higher than in the Galactic disc.  Assuming that GMCs in these violent star-forming environments remain in virial equilibrium, \citet{az01} note that the corresponding high external pressures ($P_{ext}/k \simeq 10^7-10^8\,K.cm^{-3}$) compress GMCs such that their radii become a few parsecs, similar to those of globular clusters (see solid line with filled circles in Fig.~\ref{fig:mrGMC}: $1\,pc \lesssim r_{GMC} \lesssim 10\,pc$ for $10^4\,M_{\sun} \lesssim m_{GMC} \lesssim 10^6\,M_{\sun}$).  They thus propose that compressed GMCs are the gaseous precursors of massive star clusters formed in starbursts and mergers.  If the mean surface density of GMCs in mergers and starbursts is about constant, as it is the case for Galactic GMCs, and if their internal structure resembles what is depicted by Fig.~\ref{fig:sketch}, then the mass functions of CFRgs and embedded-clusters are steeper than the GMC mass function, with the embedded-cluster index $\beta_\star$ depending on the GMC density index $p$ as shown by the (red) solid line with filled circles of Fig.~\ref{fig:betap}.     

\citet{az01} also propose that the high pressures and densities achieved in GMCs in starbursts and mergers are conducive to high SFEs.  This is in exact agreement with our scenario of denser molecular clumps (clouds) having higher global SFEs (i.e. in Eq.~\ref{eq:sfe}, a lower $\chi$ leads to a higher $\epsilon_{global}$).  
Figure~\ref{fig:mrGMC} shows the mass-radius relations of constant surface density clouds bounded by pressures of $P_{ext}/k = 5.10^6\,K.cm^{-3}$ \citep[characteristic of the Galactic centre, ][]{spe92,jog96} and $P_{ext}/k = 10^8\,K.cm^{-3}$ \citep[characteristic of galaxy mergers,][]{jog92}.  The compression of GMCs by these high external pressures raise their volume densities, rendering them closer or even similar to what is observed for Galactic CFRgs, namely, $n_{H2} \simeq 10^{4-5}cm^{-3}$.  As a result, the GMC mass fraction with $\rho \geq \rho _{th}$ increases and so does the GMC global SFE.  For $P_{ext}/k = 10^8\,K.cm^{-3}$, the normalization of the mass-radius relation is $\chi =0.008$.  Combined with $\delta=1/2$ (constant surface density), $p=1.88$ (to get $\beta_\star = 2$ with $\beta_0=1.7$, see Fig.~\ref{fig:betap} and Eq.~\ref{eq:slopes})
and a star formation density threshold $\rho_{th}=7000M_{\sun}$, this leads to a dense gas mass fraction of (Eq.~\ref{eq:mth/mcSigma}): 
\begin{equation}
\frac{m_{th}}{m_{GMC}}=6.9 \times \left(\frac{m_{GMC}}{M_{\sun}}\right)^{-0.3}\,.
\label{eq:fhd_GMC}
\end{equation}
That is, GMCs of masses $10^4\,M_{\sun}$ and $10^6\,M_{\sun}$ have $44$\,\% and $11$\,\% of their mass at a density higher than $\rho_{th}$.  With a local SFE of 35\,\%, this corresponds to global SFEs of $15$\,\% and $4$\,\%, respectively.  In comparison, the overall SFE in Galactic GMCs is of the order of $1$\,\%.  

That the mass fraction of dense gas in GMCs is a decreasing function of the GMC mass also implies that the most massive GMCs are not necessarily the largest providers of newly formed stars, despite them containing most of the molecular gas (when $\beta_0 =1.7$, clouds more massive than $10^4\,M_{\sun}$ contain $\simeq 80$\,\% of the total gas mass for a GMC mass range $10^2$~-~$10^6\,M_{\sun}$).  For the parameters adopted here, it is easy to show that each decade of GMC mass contributes an equal fraction of the total mass in dense gas.  Actually, the amount of dense gas $m_{th}^{l-u}$ contained within the GMC mass range $[m_{GMC,low},m_{GMC,up}]$ obeys $m_{th}^{l-u} \propto \ln(m_{GMC,up}/m_{GMC,low})$ and is thus constant for any given logarithmic mass interval, as expected for a CFRg mass function with $\beta=2$. \\

We emphasize that, in our scenario, the high pressures characteristic of starbursts and mergers {\it do not modify the mass-radius relation of CFRgs}.  This still equates to $n_{H2} \simeq$ a few $10^4$-$10^5cm^{-3}$, similar to CFRgs in the Milky Way disc.  What high-pressures of violent star-forming environments do modify compared to the quiesent environment of disc spirals is the mass fraction of molecular gas that is dense enough to form stars.  This mass fraction is much higher in e.g. ULIRGs than in quiescent spirals, thereby increasing galaxy star formation rates and infrared luminosities \citep{gao04}.  

\subsection{Molecular clumps as cluster-forming sites}
\label{subsec:clumps}
CO mapping (emission-lines $^{12}CO$, $^{13}CO$ or $C^{18}O$) of molecular cloud structures reveal power-law mass spectra of index $\beta_0 \simeq 1.7$, from a fraction of a solar mass up to $10^4\,M_{\sun}$ \citep{hei98, kra98, har99,won08}.  Since the same index also describes GMC data \citep{ros05,bli06,fuk08}, the mass range over which $\beta_0 \simeq 1.7$ holds covers more than 6 orders of magnitude.  

Alike to GMCs, the clump mass function is thus shallower than the young cluster mass function.  Can we correct for this effect by arguing that $CO$-mapped clumps have a constant surface density, as we have done for GMCs?  In what follows, we consider the $C^{18}O$ data of Fig.~\ref{fig:isomth} to illustrate the issue.  

The mass-radius relation of C$^{18}$O clumps is -- in essence -- one of constant {\it volume} density ($\delta = 1/3$) since their observed volume density corresponds to that needed to excite $C^{18}O$ emission (see middle panel of Fig.~\ref{fig:isomth} in Section \ref{sec:cc}).  For $\delta = 1/3$, our model predicts no steepening of the clump-to-CFRg mass functions ($\beta_0 = \beta$, Fig.~\ref{fig:beta}), and it thus seems that we cannot explain why open clusters have $\beta_\star = 2$.  However, we are interested in {\it cluster-forming} molecular clumps, and the middle panel of Fig.~\ref{fig:isomth} suggests that their mass-radius relation may actually be steeper than $\delta = 1/3$.

\begin{figure}
\includegraphics[width=\linewidth]{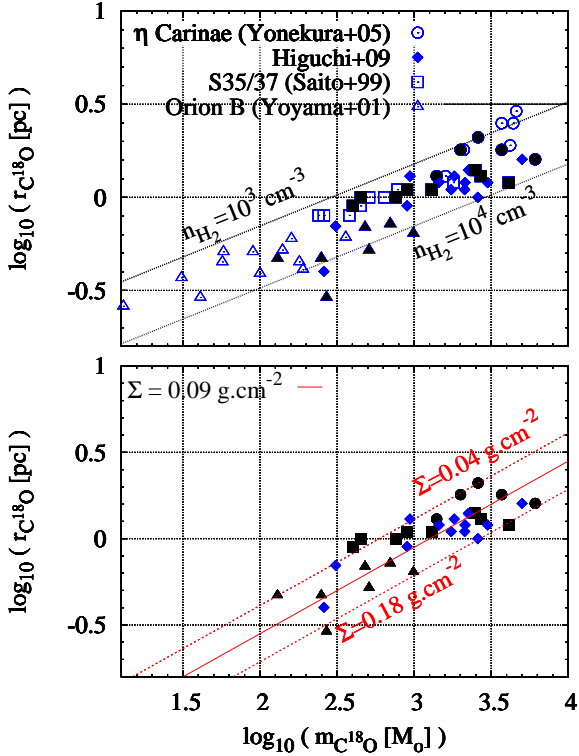}
\caption{ Radius vs. mass of $C^{18}O$ clumps.  {\it Top panel:} Filled symbols indicate detected star formation activity.  Dotted (black) lines highlight the limited range of {\it volume} densities occupied by the data (same as middle panel of Fig.~\ref{fig:isomth}, except for the added data of \citet{hig09}).  {\it Bottom panel:} Same mass-radius diagram as top panel but for star-forming $C^{18}O$ clumps only.  Dashed (red) lines indicate the limited range of {\it surface} densities occupied by the data.   \label{fig:mcSF} }
\end{figure}

The top panel of Fig.~\ref{fig:mcSF} shows the same data completed with the $C^{18}O$ clump sample of \citet[][their table 3]{hig09}.  Their 14 molecular clumps display signs of star formation activity since they were selected based on their association with an embedded-cluster.  In a follow-up study, \citet{hig10} detect $H^{13}CO^+$-line emission in all but one of them.  This strengthens the key-hypothesis of our model that star formation and high-density gas ($n_{H2} \simeq 10^5\,cm^{-3}$) are tightly related. 
Star-forming $C^{18}O$ clumps (34 in 62 objects) are highlighted by filled symbols.  For the sake of clarity, they are also shown in the bottom panel of Fig.~\ref{fig:mcSF}.    

In contrast to all $C^{18}O$ clumps, those hosting star-formation occupy a band of narrow {\it surface} density, of lower and upper limits $\Sigma _{clump} \simeq 0.04g.cm^{-2}$ and $\Sigma _{clump} \simeq 0.18g.cm^{-2}$ (dotted (red) lines), respectively.  Their mean surface density is $\Sigma _{clump} \simeq 0.09g.cm^{-2}$, or $(r_{clump}/1\,pc) \simeq 0.03 (m_{clump}/1M_{\sun})^{1/2}$.  Therefore, the mass-radius relation of cluster-forming $C^{18}O$ clumps is steeper ($\delta \simeq 1/2$) than the mass-radius relation of the whole $C^{18}O$ clump sample ($\delta \simeq 1/3$).  In turn, this steepens the CFRg mass function compared to the $C^{18}O$ clump mass function, resulting in $\beta_\star = \beta \simeq 2$ if the clump density index is $p \simeq 1.9$        

This effect can be quantified further by fitting straightlines to the $C^{18}O$ data.  We have performed robust fits \citep{pre92}, namely, fits in which the absolute value of the deviation $\Delta$ is minimized.  We consider two cases, the $y$-data being either $\log(r_{C^{18}O})$ (Eqs.~\ref{eq:fit1} and \ref{eq:fit4}) or $\log(m_{C^{18}O})$ (Eqs.~\ref{eq:fit2} and \ref{eq:fit5}).  The comparison of both fits provides a more realistic estimate of the actual uncertainties than fitting $\log(r_{C^{18}O})$ vs. $\log(m_{C^{18}O})$ (or vice-versa) alone.

Fitting all $C^{18}O$ clumps (top panel of Fig.~\ref{fig:mcSF}) gives: \\
\begin{equation} 
\log(r_{C^{18}O})=0.33\log(m_{C^{18}O})-0.96,~~ \Delta=0.08\,, 
\label{eq:fit1}
\end{equation} 
\begin{equation}
\log(m_{C^{18}O})=2.41\log(r_{C^{18}O})+2.90,~~ \Delta=0.21\;,
\label{eq:fit2}
\end{equation}
the latter equating with:
\begin{equation}
\log(r_{C^{18}O})=0.41\log(m_{C^{18}O})-1.20\;,
\label{eq:fit3}
\end{equation}
Taking the mean slopes and intercepts of Eqs.~\ref{eq:fit1} and \ref{eq:fit3} gives:
\begin{equation}
\log(r_{C^{18}O})=0.37\log(m_{C^{18}O})-1.08\,. 
\label{eq:fitall}
\end{equation}
Performing the same robust fits onto $C^{18}O$ clumps hosting forming-clusters (bottom panel of Fig.~\ref{fig:mcSF}) provides: \\
\begin{equation} 
\log(r_{C^{18}O})=0.36\log(m_{C^{18}O})-1.08,~~ \Delta=0.08\,, 
\label{eq:fit4}
\end{equation} 
\begin{equation}
\log(m_{C^{18}O})=1.92\log(r_{C^{18}O})+3.03,~~ \Delta=0.18\,,
\label{eq:fit5}
\end{equation}
the second equation being equivalent to:
\begin{equation}
\log(r_{C^{18}O})=0.52\log(m_{C^{18}O})-1.58 \,.
\label{eq:fit6}
\end{equation}
Averaging slopes and intercepts of Eqs.~\ref{eq:fit4} and \ref{eq:fit6} leads to:
\begin{equation}
\log(r_{C^{18}O})=0.44\log(m_{C^{18}O})-1.33\,. 
\label{eq:fitSF}
\end{equation}
Excluding clumps failing at displaying evidence of star formation indeed steepens the clump mass-radius relation, although the effect is mild (compare Eqs.~\ref{eq:fitall} and \ref{eq:fitSF}).  It is mostly driven by the exclusion of the low-mass clumps in Orion~B ($m_{C^{18}O}<150\,M_{\sun}$) whose undetected (non-existent~?) star-formation activity may stem from a dearth of high-density ($n_{th} \geq 10^5\,cm^{-3}$) molecular gas, as discussed at the end of Section \ref{subsec:mod}.  A better handling of the mass-radius relation of cluster-forming $C^{18}O$ clumps, compared to that of $C^{18}O$ clumps in general, would require data covering a larger mass range.   \\

Conversely, one may expect cluster-forming molecular clumps with an observed mass function $\beta_0=2$ to be in a regime of constant mean volume density (since constant volume density does not alter the mass function slope, i.e. $\beta = \beta_0$ when $\delta=1/3$; see Fig.~\ref{fig:beta}).  Several dust continuum studies report clump mass functions with indices $\beta_0 \simeq 2$.  However, few of them are characterized by a significant number of clumps more massive than, say, $100\,M_{\odot}$, that is, a mass regime appropriate for star cluster progenitors rather than individual star progenitors.  

\citet{rat06} find a mass function slope of $-2.1\pm0.4$ for a sample of dust clumps with masses $>100\,M_{\odot}$ and mapped in millimeter continuum.  The corresponding clump mass-diameter diagram (their fig.~8) shows a significant scatter and no clear-cut mass-size relation.  That the vast majority of clumps are denser than $10^4\,cm^{-3}$ is the only firm conclusion one can reach.  More studies of that type are needed before drawing a conclusion.  We encourage authors of such surveys to publish the clump radius distribution and clump mass-radius diagram in addition to the clump mass function, especially if the clump mass upper limit reaches several $10^3\,M_{\odot}$ and beyond.  This implies to include star-forming regions more distant from the Sun than a few kpc.  In that respect, it should be kept in mind that clump mass and radius estimates depend on the assumed clump distance $D$ ($r_{clump} \propto D$ through the clump angular diameter; $m_{clump} \propto D^2$, see eq.~1 in \citet{rat06}).  It may be of interest to test how the scatter of a mass-radius diagram responds to varying the clump distance accuracy.  

Finally, we note for the sake of completeness that the existence of a link between the slope of the clump mass function and the clump volume density was put forward by \citet{rw05}, who quote that "a possible explanation for the apparently real discrepancy between the CO spectral line and dust continuum mass functions is that the dust maps trace denser clumps than the CO line maps".  [But see \citet{mun07} for a counter-argument following which steep mass functions inferred by some dust continuum studies are an artifact created by the clump mass upper limit.]

\section{From the clump radius distribution to the cluster-forming region radius distribution}
\label{sec:rdist}

Similarly to what we have done in Section \ref{subsec:mf} to relate the clump and CFRg mass functions, we now infer the radius distribution of CFRgs from that of their parent clumps.  

Let us consider a population of clumps whose radius distribution is a power-law of slope $-x_0$:
\begin{equation}
dN = l_{clump} r_{clump}^{-x_0} dr_{clump}\;.
\label{eq:cordist}
\end{equation}
If clump masses and radii are correlated (i.e. $\delta \ne 0$ in Eq.~\ref{eq:rcmc}), then the slope $-x_0$ of the radius distribution of clumps is determined by the slope $\delta$ of their mass-radius relation and the slope $-\beta_0$ of their mass function.  To show that, we use Eq.~\ref{eq:rcmc} to replace $r_{clump}$ as a function of $m_{clump}$ and $\delta$ in Eq.~\ref{eq:cordist}.  This leads to the clump mass function with its slope $-\beta_0$ a function of $\delta$ and $x_0$:

\begin{equation}
dN \propto r_{clump}^{-x_0} dr_{clump} \propto m_{clump}^{-[1+\delta(x_0-1)]} dm_{clump}\,.
\label{eq:rdist}
\end{equation}
It then follows:
\begin{equation}
\delta = \frac{\beta_0-1}{x_0-1}\,.
\label{eq:debex}
\end{equation}  
To derive the radius distribution of the dense central CFRgs: 
\begin{equation}
dN = l_{th} r_{th}^{-x} dr_{th}\,,
\label{eq:clrdist}
\end{equation}
we need to derive the CFRg radius $r_{th}$ as a function of the clump radius $r_{clump}$.
By combining Eqs~\ref{eq:rcmc} and \ref{eq:rth}, we obtain: 
\begin{equation}
r_{th}=\left(\frac{3-p}{4\pi \rho_{th} \chi^{1/\delta}}\right)^{1/p} r_{clump}^{[(p-3)\delta+1]/(\delta p)} 
\label{eq:rthbis}
\end{equation}
where $\chi$ and $\delta$ are the normalization and slope of the clump mass-radius relation (Eq.~\ref{eq:rcmc}), $p$ is the clump density index (Eq.~\ref{eq:rho}), and $\rho_{th}$ is the volume density at the edge of the CFRg.
Combining Eq.~\ref{eq:cordist} and Eq.~\ref{eq:rthbis} leads to the CFRg radius distribution:
\begin{eqnarray}
dN=l_{clump} \left(\frac{4 \pi \rho_{th} \chi^{1/\delta}}{3-p}\right)^{[\delta(1-x)]/[(p-3)\delta+1]} \nonumber \\
\frac{p\delta}{(p-3)\delta+1} r_{th}^{-x} dr_{th},
\label{eq:clrdist1}
\end{eqnarray}
where the slope $-x$ obeys:
\begin{equation}
-x=-\frac{(x_0 p-3)\delta+1}{(p-3)\delta+1}\,.
\label{eq:x1}
\end{equation}

\begin{figure}
\includegraphics[width=\linewidth]{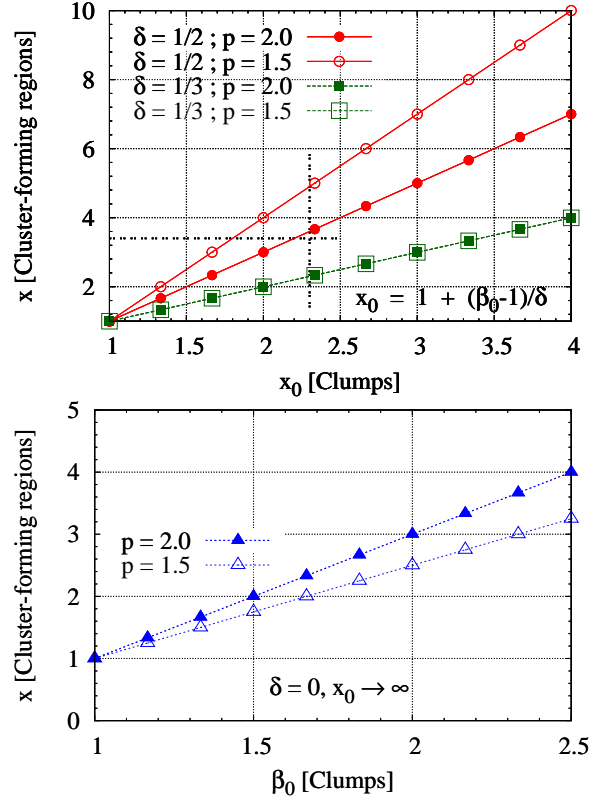}
\caption{ {\it Top panel:} Slope $x$ of the cluster-forming region (CFRg) radius distribution in dependence of the slope $x_0$ of the clump radius distribution for constant surface density ($\delta =1/2$) and constant volume density ($\delta =1/3$) clumps.  Two density profiles ($p=2$ and $p=1.5$) are envisaged.  Note that the choice of $x_0$ and $\delta$ necessarily determines the spectral index $\beta_0$ of the clump mass function through Eq.~\ref{eq:debex}.  The vertical (black) dashed line indicates the clump radius distribution slope when $\delta =1/2$ and $\beta_0=1.7$.  {\it Bottom panel:} How $x$ scales against the clump mass function spectral index $\beta_0$ when $\delta =0$ (i.e. clump radius independent of clump mass).  Note that $\delta=0$ implies $x_0 \rightarrow \infty$.    \label{fig:x} }
\end{figure}

In the particular case $\delta =0$ in Eq.~\ref{eq:rcmc} (i.e. clump masses and radii are uncorrelated), Eq.~\ref{eq:debex} shows that the slope of the radius distribution $-x_0 \rightarrow -\infty$.  Therefore, Eq.~\ref{eq:cordist} cannot be used to infer the CFRg radius distribution.  Using the clump mass function (Eq.~\ref{eq:coMF}) and the CFRg radius as a function of $m_{clump}$ (Eq.~\ref{eq:rth}) instead, one obtains the distribution function of the CFRg radius $r_{th}$, with the clump radius $r_{clump}$ a constant:
\begin{eqnarray}
\label{eq:clrdist2}
dN=k_{clump} \, p \, \left( \frac{4\pi \rho_{th}}{3-p} r_{clump}^{3-p} \right)^{1-\beta} \nonumber \\
r_{th}^{-x} dr_{th}\,.
\end{eqnarray} 
where the slope $-x$ obeys:
\begin{equation}
-x=-[1+p(\beta_0 -1)]\,.
\label{eq:x2}
\end{equation}

\begin{figure}
\includegraphics[width=\linewidth]{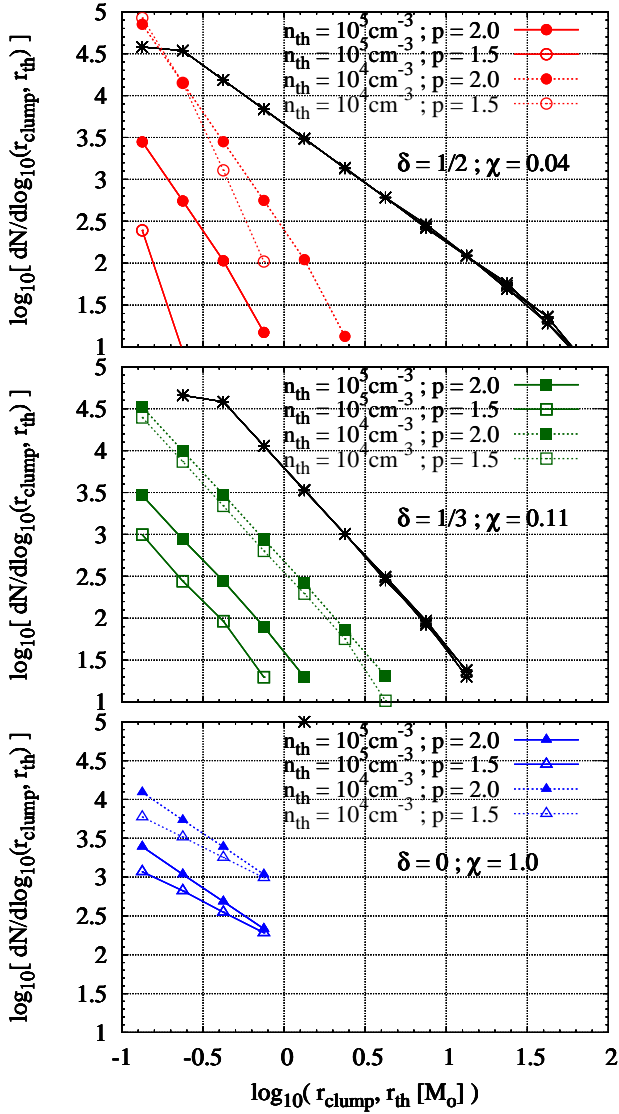}
\caption{Comparison between the clump (solid black lines with asterisks) and CFRg radius distributions, for clumps with constant surface density, constant volume density and constant radius (top, middle and bottom panels, respectively).  The radius distribution of clumps is built from their mass function used in Fig.~\ref{fig:MF} ($\beta_0 = 1.7$) and a given clump mass-radius relation ($\chi$, $\delta$, see Eq.~\ref{eq:rcmc}).  Symbol- and colour-codings are identical to Fig.~\ref{fig:beta}.     \label{fig:rdist} }
\end{figure}

Equations \ref{eq:x1} and \ref{eq:x2} are shown in the top ($\delta=1/3$, $\delta=1/2$) and bottom ($\delta=0$) panels of Fig.~\ref{fig:x}.  The same density indices ($p=2$ and $p=1.5$) and colour/symbol-codings as previously are used.  As for the mass function, $\delta=1/3$ leads to identical slopes of the clump and CFRg radius distributions, while $\delta=1/2$ and $\delta=0$ result in CFRg radius distributions steeper and shallower than the clump radius distribution, respectively.  Note that in the bottom panel the index $x$ of the CFRg radius distribution is plotted as a function of the clump mass function index $\beta_0$ since $x_0 \rightarrow \infty$.    \\

In Fig.~\ref{fig:rdist}, we illustrate how the radius distribution of CFRgs differs from that of their host-clumps (solid black lines with asterisks), for clumps with constant mean surface density, mean volume density and radius (top, middle and bottom panels, respectively).  Each panel displays 4 cases corresponding to two clump density indices ($p=2$ and $p=1.5$) and two number density thresholds $n_{th}$ to define the CFRg ($n_{th}=10^4\,cm^{-3}$ and $n_{th}=10^5\,cm^{-3}$).  These simulations are the counterparts of those performed in Section \ref{subsec:mf} to study the transition from the clump- to CFRg-mass functions.  Lower densities $n_{th}$ at the edge of the CFRg result in smaller shifts between the clump and CFRg radius distributions since CFRgs are then larger in size. \\

Assuming as in Section \ref{subsec:GMC} that GMCs host the precursors of star clusters, can these power-law models of clump and CFRg radius distributions help us understand the radius distributions of clusters and GMCs?  \citet{sss85} find that the distribution of GMC diameters $D$ obeys $N(D) \propto D^{-2.3\pm0.25}$ in our Galaxy, i.e. $x_0=2.3$ in Eq.~\ref{eq:cordist}.  This slope, indicated as the (black) dashed vertical line in top panel of Fig.~\ref{fig:x}, is in excellent agreement with what Eq.~\ref{eq:debex} predicts for constant mean surface density clouds ($\delta = 1/2$) with $\beta_0 \simeq 1.7$, that is, $x_0=2.4$.  As for star clusters, their size is often embodied by their half-light radius $r_{hl}$.  It appears that the distribution function of cluster half-light radii is not fully comparable to that of GMC diameters.  While the Galactic GMC size distribution is a power-law, observed distributions of cluster half-light radii are characterised by an intrinsic peak when plotted as $dN/d\log(r_{hl})$, that is, as the number of clusters per constant logarithmic radius interval \citep{az98}.  More recently, \citet{sch07} also highlighted such a shape for the young star clusters of the Whirlpool galaxy M51 (see their fig.~14).   Their $dN/d\log(r_{hl})$ distribution shows a turnover at $r_{hl} \simeq 2$\,pc, while the distribution increasingly steepens beyond $r_{hl} > 4$\,pc.  Therefore, the comparison of the half-light radius distributions of clusters to the size distribution of GMCs is necessarily limited to the large cluster-radius regime (say, $r_{hl} > 4$\,pc), a point also made by \citet{az01}.  In that regime, \citet{az01} report an index  $x \simeq 3.4$ for the half-light radius distributions of old Galactic globular clusters and young massive star clusters in NGC3256.  That is, the radius distribution of star clusters is steeper than that of GMCs.  This $x$ value is shown as the (black) dashed horizontal line in top panel of Fig.~\ref{fig:x}.  Figure~\ref{fig:x} demonstrates that our model does account for this effect since constant surface density clouds -- as is observed for GMCs -- steepens the radius distribution over the cloud-to-CFRg transition ($x > x_0$).  As for the mass function, the shallower the cloud density index, the stronger the radius distribution steepening.  In the case of relevance here, the observed $x-x_0$ difference is reproduced when the density index is $p \simeq 2$ (see the solid red line with filled circles in top panel of Fig.~\ref{fig:x}).

It must be kept in mind, however, that to identify the slope $-x$ of the CFRg radius distribution to the slope of observed cluster half-light radius distributions may constitute a severe oversimplification, even when the comparison is restricted to the large-radius regime.  Following gas-expulsion, embedded-clusters expand \citep[][their fig.~3]{gey01} and may then undergo tidal truncation \citep{par10}, two effects which may complicate the picture significantly.  More simulations covering the evolution from the gas-embedded phase to the end of violent relaxation are required before drawing definitive conclusions.      

\section{Conclusions}
\label{sec:conclu}

It has long been recognized that the mass function $dN \propto m^{-\beta_0} dm$ of GMCs and of their molecular clumps mapped in CO-emission line is shallower than the `canonical' young cluster mass function $dN \propto m^{-\beta_\star} dm$, i.e. $\beta_0 \simeq 1.7$ and $\beta_\star \simeq 2$.  This slope difference is puzzling since it seemingly implies an SFE varying with the GMC or clump mass, hence mass-dependent cluster infant weight-loss while the cluster responds to gas-expulsion.  This is in contradiction with most young cluster mass function data gathered so far. 

In this contribution we bring an original solution to this problem by assuming that star formation requires a {\it number} density threshold $n_{th} \simeq 10^{4-5}\,cm^{-3}$, equivalent to a {\it volume} density threshold $\rho_{th} \simeq 700-7000\,M_{\sun}.pc^{-3}$.  This hypothesis is supported by the tight association observed between star-formation and dense molecular gas (as evidenced by e.g. $H^{13}CO^+$ and $HCN$ tracers; see Section \ref{sec:evid}).  Our model builds on a spherically symmetric cloud (or clump) with a power-law density profile and forming a star cluster in its central region.  The density threshold for star formation $\rho_{th}$ is not necessarily achieved through the whole molecular cloud (clump), thereby implying that the mass and radius of the CFRg differ from those of the cloud (or clump) containing it.  We refer to $\beta$ as the mass function index of the spatially-limited CFRg where $n_{H2} \geq n_{th}$ (see Fig.~\ref{fig:sketch}).  

In that context, star formation can be quantified by two distinct efficiencies {\it of different physical significances}.  We refer to the {\it global} SFE as the ratio between the embedded-cluster stellar mass at the onset of gas-expulsion and the initial gas mass of the {\it clump (cloud)} hosting it.  As such, the global SFE is relevant to understand the difference between the cloud (clump) mass function on the one hand, and the embedded-cluster mass function on the other hand.  In contrast, the {\it local} SFE quantifies the ratio between the embedded-cluster stellar mass and the initial gas mass of the CFRg, i.e. the gas mass with $n_{H2} \geq n_{th}$.  This is the local SFE -- {\it not} the global one -- which is relevant to understand why cluster violent relaxation is mass-independent.  Mass-independent infant weight-loss demonstrates that the local SFE is CFRg-mass-{\it in}dependent hence that the slopes of the CFRg and embedded-cluster mass functions are identical ($\beta = \beta_\star$).  This does not prevent the global SFE from being {\it clump/cloud} mass-dependent, as suggested by the difference in slope $\beta_\star - \beta_0$ between the cloud (clump) and cluster mass functions.  

To adopt a volume density threshold for cluster formation immediately implies that CFRgs have a constant mean volume density (Eq.~\ref{eq:av_rhoth}).  Based on the conditions required for the tidal-field impact upon clusters responding to gas-expulsion to be mass-independent, this is also the conclusion reached by \citet{par10}.   Actually, not only does mass-independent violent relaxation demand mass-independent local SFE, it also requires mass-independent gas-expulsion time-scale \citep{par08b}, and mass-independent tidal-field impact \citep{par10}.     

We have shown that the difference in slope between the clump- (cloud-) and CFRg-mass functions is a sensitive function of the mass-radius relation and density index $p$ of clumps (clouds).  Constant radius clumps result in the mass function of CFRgs (hence of embedded-clusters) being shallower than the mass function of their host clumps (clouds).  This is due to more massive clumps being denser, thus containing a greater fraction of their mass above the number density threshold $n_{th}$.  Equivalently, the global SFE increases with the clump mass.  Conversely, 
the volume density of constant surface density clumps is a decreasing function of their mass, and so is their mass fraction of star-forming gas.  This renders the mass function of CFRgs steeper than that of clumps/clouds ($\beta > \beta_0$).  For constant volume density clumps/clouds, CFRg and cloud/clump mass function slopes are alike (Fig.~\ref{fig:beta}).  Given a cloud (clump) mass-radius relation, the slope difference $|\beta-\beta_0|$ depends on the density index $p$ of clumps (clouds): the shallower the clump/cloud density profile, the larger $|\beta-\beta_0|$ (Fig.~\ref{fig:betap}).  

The steepening of the mass function $\beta_0 \simeq 1.7$ of molecular clumps and GMCs into that $\beta_{\star} \simeq 2$ of young star clusters therefore requires molecular clouds and clumps to have a constant surface density (Fig.~\ref{fig:MF}).  This property is actually well-established for GMCs in the Milky Way \citep{bli06, hey09}.  Whether it also stands for molecular clumps, namely, the density enhancements -- birth sites of open clusters -- observed locally within Galactic GMCs,  is less certain.  Rather, molecular clumps show a constant volume density corresponding to that required to excite the molecular transition of relevance (middle panel of Fig.~\ref{fig:isomth}).  Based on $C^{18}O$ data, we speculate that the mass-radius relation of {\it cluster-forming} molecular clumps is one of constant surface density, rather than of constant volume density (see Fig.~\ref{fig:mcSF}).  The transition from a narrow range in volume densities for all $C^{18}O$ clumps to a narrow range in surface densities for those with signs of star formation stems from excluding the lowest mass clumps.  Those clumps contain a tiny mass only with $n_{H2} \geq n_{th}$, which  explains their failure at displaying signs of star formation (top panel of Fig.~\ref{fig:isomth}).  

From their survey in dust-continuum emission of star-forming regions in the Galactic disc, \citet{mue02} infer a mean density index $p \simeq 1.8$.  Interestingly, in that case, we find that the mass function slope $\beta_0 \simeq 1.7$ of clouds (clumps) steepens into a CFRg mass function slope $\beta \simeq 2$ (Fig.~\ref{fig:betap}) hence $\beta_{\star} \simeq 2$, in agreement with what is suggested by observations.  Equivalently, the global SFE of molecular clouds (clumps) is a decreasing function of their mass (Fig.~\ref{fig:sfe}).      

A natural outcome of our model is that as mapping of molecular clumps move inwards to their higher-density  CFRgs, the inferred mass function is expected to steepen and to near $\beta \simeq 2$.  This may be the reason why \citet{shi03} find $\beta \simeq 1.9$ for CFRgs mapped in $CS (J~5 - 4)$ which, as they quote, is steeper than what is measured with tracers of lower density gas, and closer to the mass spectral index of OB associations (their fig.~20).     

In addition to the mass functions, we have also studied the radius distributions.  Given constant mean surface density clouds (clumps), not only does our model steepen the mass function, it also steepens the radius distribution (top panel of Fig.~\ref{fig:rdist}).  The slope of the radius distribution of clouds (clumps) is determined by their mass-radius relation and mass function slopes (Eq.~\ref{eq:debex}).  Constant surface density clouds (clumps) ($\delta=1/2$) with $\beta_0 = 1.7$ have a radius distribution $dN \propto r^{-x_0} dr_{clump}$ of index $x_0=2.4$.  For a density index $p=1.8$, the slope of the radius distribution steepens to $x \simeq 4$ for CFRgs hence embedded-clusters (top panel of Fig.~\ref{fig:x}).  Our model thus helps explain why the cluster radius distribution is significantly steeper than the size distribution of GMCs \citep{sss85,sch07}.

\section{Acknowledgments}
G.P. acknowledges support from the Alexander von Humboldt Foundation in the form of a Research Fellowship, and from the Humboldt-Professorship of Prof.~Norbert Langer.  G.P. is also grateful to Uta Fritze for past lively discussions which have proven most helpful on the long-term.

\end{document}